\documentclass[twocolumn]{revtex4}

\usepackage{graphicx}
\usepackage{bm}
\usepackage{amsmath}
\usepackage{amssymb}
\usepackage{rotating}

\begin{document}
\title{Enhanced sampling in generalized ensemble with large gap of sampling parameter:
case study in temperature space random walk}

\author{Cheng~Zhang} %
\affiliation{Department of Bioengineering, Rice University, Houston,
Texas 77005, USA}

\author{Jianpeng~Ma}
\email{jpma@bcm.tmc.edu} %
\affiliation{Department of Bioengineering,
Rice University, Houston, Texas 77005, USA} %
\affiliation{ Verna and Marrs McLean Department of Biochemistry and
Molecular Biology, Baylor College of Medicine, Houston, Texas 77030,
USA }

\date{\today}

\begin{abstract}
We present an efficient sampling method for computing a partition
function  and accelerating configuration sampling.
The method performs a random walk in the $\lambda$ space, with
$\lambda$ being any thermodynamic variable that characterizes a
canonical ensemble such as the reciprocal temperature $\beta$ or any
variable that the Hamiltonian explicitly depends on.
The partition function is determined by minimizing the difference of
the thermal conjugates of $\lambda$ (the energy in the case of
$\lambda=\beta$), defined as the difference between the value from
the dynamically updated derivatives of the partition function and
the value directly measured from simulation.
Higher-order derivatives of the partition function are included to
enhance the Brownian motion in the $\lambda$ space.
The method is much less sensitive to the system size, and the size
of $\lambda$ window than other methods.
On the two dimensional Ising model, it is shown that the method
asymptotically converges the partition function, and the error of
the logarithm of the partition function is much smaller than the
algorithm using the Wang-Landau recursive scheme. The method is also
applied to off-lattice model proteins, the $AB$ models, in which
cases many low energy states are found in different models.
\end{abstract}

\maketitle

\newcommand{\Hr}{\mathcal H_\lambda}
\newcommand{\Avl}{\Bigl\langle}
\newcommand{\Avr}{\Bigr\rangle}
\newcommand{\avl}{\bigl\langle}
\newcommand{\avr}{\bigr\rangle}
\newcommand{\wavr}{\avr_{\hspace{-.2ex}\tilde w}}
\newcommand{\wAvr}{\Avr_{\hspace{-.2ex}\tilde w}}
\newcommand{\lavr}{\avr_{\hspace{-.2ex}\lambda}}
\newcommand{\lAvr}{\Avr_{\hspace{-.2ex}\lambda}}

\section{Introduction}
In a canonical ensemble, the partition function is defined as a sum
over configurations (denoted by $X$),
\begin{equation} %
Z(\lambda) = \sum_X \exp\bigl[-\Hr(X)\bigr],%
\label{partitionfunction}
\end{equation}
where $\Hr(X)$ is the reduced Hamiltonian of the system with a
dependence on a variable $\lambda$.  In the case that the $\lambda$
dependence only exists in the energy function $E_\lambda(X)$,
$\Hr(X)$ can be written as $\beta E_\lambda(X)$, where $\beta=1/T$
is the reciprocal temperature. However,  $\lambda$ can be the
temperature $\beta$, leaving the energy function $E(X)$ independent
of $\beta$.

A quantity of particular interest is the ratio of the partition
function $Z(\lambda_1)/Z(\lambda_0)$ at two given $\lambda$'s,
$\lambda_1$ and $\lambda_0$ (if $\lambda$ does not explicitly
involve the temperature, the ratio can be translated to the free
energy difference as $ \Delta F = - T
\ln\bigl[Z(\lambda_1)/Z(\lambda_0)\bigr]$).
In our previous study \cite{Z}, the partition function is computed
in an expanded ensemble, where a regular simulation is coupled with
random transitions among different $\lambda$'s, e.g., temperatures
$\beta$'s or volumes $V$'s. This approach requires two neighboring
distributions of the corresponding macroscopic quantities, such as
the energy for the temperature or the virial for the volume, to
overlap sufficiently. Accordingly the spacing $\Delta\lambda$ is
proportional to $1/\sqrt N$, and the number of $\lambda$ sampling
points increases as $\sqrt N$, as the system size $N$ grows.
Many other methods, such as replica exchange \cite{replica},
simulated tempering \cite{simtemp}, and others \cite{adaptive,
sscaling}, has similar issures.

To overcome this problem of increasingly large number of sampling
points, we define $\lambda$ as a continuous variable instead of a
discrete one.
The partition function $Z(\lambda)$ is characterized as a continuous
function by a few adjustable parameters (e.g., derivatives with
respect to $\lambda$) for a large $\lambda$ window.
In this way, the number of $\lambda$ windows can be significantly
reduced, and one can handle a much larger system.
Moreover, the ratio of the partition function at endpoints (i.e.,
boundaries of a $\lambda$ window), $Z(\lambda_1)/Z(\lambda_0)$, can
be asymptotically determined through a simulation.

In this paper, we first present the theoretical background of the
method in a general framework. Next the simulation protocol is
exemplified in a special case where $\lambda$ is the reciprocal
temperature $\beta$. We then numerically test the performance of the
method on the two-dimensional Ising model, and apply the method to
folding model proteins. At the end, we conclude the method with
discussions.

\section{Method}

\subsection{General Theory}
We start by constructing a generalized ensemble composed of
canonical ensembles of a continuous  $\lambda$ range. The
probability of $\lambda$ being in the interval $(\lambda,
\lambda+d\lambda)$ is given by
\begin{equation}
\begin{split}
w(\lambda)d\lambda %
&= \tilde w(\lambda) \frac{Z(\lambda)}{\tilde Z(\lambda)} \ d\lambda \\ %
&= \sum_X \tilde w(\lambda) \exp\Bigl[ - \Hr(X) -\ln\tilde Z(\lambda)\Bigr] %
  d\lambda, %
\label{wZ} %
\end{split}
\end{equation}
where we have introduced an approximate partition function $\tilde
Z(\lambda)$, as well as a predefined weight function $\tilde
w(\lambda)$.  In the second line, the partition function is expanded
using Eq. (\ref{partitionfunction}).  If, in a special case, $\tilde
w(\lambda)$ is a constant, the generalized ensemble corresponds to a
flat $\lambda$ histogram \cite{Z, replica, sscaling, wl, adaptive}.

An efficient sampling has three requirements. First, the method
should yield the correct partition function ratio at the end points
$Z(\lambda_1)/Z(\lambda_0)$, or the free energy difference. Second,
the difference between the actual weight $w(\lambda)$ should be
close to the desired one $\tilde w(\lambda)$, or equivalently
$\tilde Z(\lambda)$ should be close to $Z(\lambda)$. Last, an
efficient scheme for the $\lambda$-space random walk is required.

The first requirement can be rephrased to a condition on the average
properties of the ensemble,
\begin{equation}
\begin{split}
0   & = \frac{Z(\lambda_1)}{\tilde Z(\lambda_1)}
      - \frac{Z(\lambda_0)}{\tilde Z(\lambda_0)} \\
    & = \int_{\lambda_0}^{\lambda_1} %
    \frac{\partial }{\partial \lambda} \biggl( \frac{Z}{\tilde Z}\biggr) ~d\lambda \\
    & = \int_{\lambda_0}^{\lambda_1} %
        \biggl[-\frac{\partial \ln \tilde Z}{\partial \lambda}
        -\Avl \frac{\partial \Hr}{\partial \lambda} \lAvr \biggr]
        \biggl( \frac{Z}{\tilde Z}\biggr) ~d\lambda \\
    & =  %
    \Avl \tilde P(\lambda) \wAvr -
    \Avl \ \avl P_\lambda(X)\lavr \  \wAvr.
\end{split}
\label{convergence}
\end{equation}
Here,
$\avl \ldots \lavr$ denotes a configuration average at a fixed
$\lambda$;
$\tilde P(\lambda) \equiv -\partial \ln \tilde Z/\partial \lambda$ \
is the derivative of the estimated partition function;
$P_\lambda(X) \equiv \partial \Hr(X)/\partial \lambda$ is the
thermal conjugate of $\lambda$;
$\avl A \wavr$ denotes a weighted average in the generalized
ensemble for quantity $A$,
\[
\avl A \wavr \equiv \int_{\lambda_0}^{\lambda_1}
A(\lambda) \ \frac{w(\lambda)}{\tilde w( \lambda)} ~ d\lambda %
= \int_{\lambda_0}^{\lambda_1} A(\lambda) \cdot \frac{Z(\lambda)}{
\tilde Z(\lambda)} ~ d\lambda.
\]
It is evident that as long as the two averages on the right hand
side of the last line of Eq. (\ref{convergence}) are equal, the
requirement on the correct ratio of the partition function is
satisfied. In simulation, $\avl \tilde P(\lambda) \lavr$ from the
estimated partition function is dynamically adjusted to be equal to
the measured average $\Avl\ \avl P_\lambda(X)\lavr \ \wAvr$
(practically, the two-fold average is translated to the average in
the generalized ensemble, which is then replaced by a trajectory
average measured along simulation). As a result, one can eventually
obtain the correct ratio of the partition function.

The second requirement is satisfied by minimizing the following
quantity,
\begin{equation}
S = \Avl \Bigl[ \tilde P(\lambda)- \avl P_\lambda(X)\lavr \Bigr]^2 \wAvr. %
\label{action}
\end{equation}
Since $\tilde P(\lambda)$ and $\avl P_\lambda(X)\lavr$ are the
derivatives of $-\ln \tilde Z(\lambda)$ and  $-\ln Z(\lambda)$
respectively, $S$ is minimized when the two are equal at any
$\lambda$. %
However, a perfect match between $\tilde P(\lambda)$ and $\avl
P_\lambda(X)\lavr$ is practically impossible to reach.  We thus
adopt a variational approach, where $\tilde P(\lambda)$ is
approximated as a linear combination of a few trial functions
$\phi_k(\lambda)$'s as
\[
\tilde P(\lambda)=\sum_k a_k \phi_k(\lambda).
\]
An example of the expansion is a power series $\tilde P(\lambda)=a_0
+ a_1 \lambda + a_2 \lambda^2+ \ldots$, where, $\phi_k(\lambda) =
\lambda^k$, and $a_k$ corresponds to the $k$th order derivative of
$\tilde P(\lambda)$, $a_k = (1/k!) \hspace{0.4ex}
d^{\hspace{0.2ex}(k)} \tilde P/ d \lambda^{(k)}$.
Now one can minimize $S$ with respect to the coefficients $a_k$'s as
$\partial S/\partial a_k=0$. This determines $a_k$'s from the
following set of equations,
\begin{equation}
    \sum_k  a_k \ \Avl  \phi_j(\lambda) \  \phi_k(\lambda) \wAvr%
=   \Avl  \phi_j(\lambda) \  \avl P_\lambda(X)\lavr  \ \wAvr. %
\label{matrix}
\end{equation}
Similar to the case of Eq. (\ref{convergence}), the two-fold average
$\Avl \phi_j(\lambda) \  \avl P_\lambda(X)\lavr \ \wAvr$ is
equivalent to the average of $\phi_j(\lambda)  P_\lambda(X)$ in the
generalized ensemble, and can be evaluated from a simulation
trajectory.
The parameters $a_k$'s are regularly updated in simulation
accordingly to Eq.~(\ref{matrix}) to enforce the minimization of
$S$.
Once all $a_k$'s are obtained, the ratio of the estimated partition
function can be calculated as
\begin{equation}
    \ln \tilde Z(\lambda_1)-\ln \tilde Z(\lambda_0) %
    = - \int_{\lambda_0}^{\lambda_1} %
        \tilde P(\lambda) \ d \lambda
    =- \sum_k a_k ~\Delta \Phi_k.
    \label{lnZ}
\end{equation}
where $\Delta \Phi_k \equiv \int_{\lambda_0}^{\lambda_1}
\phi_k(\lambda') d\lambda'$.
We now show that Eq. (\ref{matrix})  is compatible with the first
requirement Eq. (\ref{convergence}):
assuming $\phi_0=1$, the first
equation of Eq. (\ref{matrix}), i.e. the $j=0$ case,
becomes $\avl \tilde P(\lambda) \wavr = \Avl \sum_k  a_k  \phi_k\wAvr%
= \Avl \ \avl P_\lambda(X)\lavr \wAvr$, which is identical to Eq.
(\ref{convergence}).

Last, sampling in the generalized ensemble can be implemented by a
regular configurational sampling at a fixed $\lambda$, as well as a
random walk in the $\lambda$ space. Any constant temperature
algorithm can be used to generate configurational moves at a fixed
$\lambda$. For the $\lambda$-space sampling, a convenient choice is
to follow a Langevin equation:
\begin{equation}
\begin{split}
    \frac{d\lambda}{dt}
    & = -\frac{1}{\tilde w}\  \Bigl[P_\lambda(X) -\tilde P(\lambda)\Bigr]  %
        +\frac{\xi}{\sqrt{\tilde w}}, %
\label{lang}
\end{split}
\end{equation}
where $\xi$ is a Gaussian white noise that satisfies  $\avl \xi(t)
\xi(t') \avr = 2 \delta(t-t')$, with $t$ being the simulation time.
The equation is derived by treating  $V_\lambda (X) = \Hr(X) + \ln
\tilde Z(\lambda)$ in Eq. (\ref{wZ}) as the ``potential'', and its
derivative $-\ \partial V_\lambda (X)/\partial \lambda = -
\Bigl[P_\lambda(X) -\tilde P(\lambda)\Bigr]$ as the ``force'' of the
$\lambda$-space random walk.
To show that Eq. (\ref{lang}) yields the correct $\lambda$
distribution, we examine the time evolution of the $\lambda$
distribution $\rho(\lambda)$, described
by the corresponding Fokker-Planck equation, %
    $\partial \rho/\partial t %
    = (\partial/\partial\lambda) %
        \bigl[(\partial V_\lambda/\partial \lambda)\hspace{.5ex} (\rho/\tilde w)\bigr] %
    + (\partial^2/\partial\lambda^2) \bigl(\rho/\tilde w\bigr) %
    $,
whose stationary solution ($\partial \rho/\partial t=0$) indeed
gives the desired distribution $\rho \sim \tilde w
~\exp\bigl[-V_\lambda (X)\bigr]$.

\subsection{A case study: temperature space sampling}

In the following discussion, $\lambda$ is assumed to be the
reciprocal temperature $\beta$.  Thus, $P_\lambda(X) =
\partial \bigl[\beta E(X)\bigr]/\partial \beta$ in Eq. (\ref{convergence}) is the energy $E$,
and $\tilde P(\lambda)$ can be interpreted as the estimated average
energy $\tilde E(\beta) = -\partial \ln \tilde Z /\partial \beta$.

The aim is to compute the ratio of the partition function
$Z(\beta_1)/Z(\beta_0)$. %
The ratio is to be calculated from the estimated average energy
$\tilde E(\beta)$.
For this reason, we look for a best fit between estimated average
energy $\tilde E(\beta)$ and the actual one $\avl E\avr_\beta$. If
one expands the estimated average energy as $\tilde E(\beta) = a_0 +
a_1\beta + a_2 \beta^2+ \ldots$, the task is to determine the best
fitting coefficients $a_k$'s.
As in the general formalism, the trial functions are $\phi_0=1,
\phi_1=\beta, \phi_2=\beta^2, \ldots$, and the coefficients
correspond to derivatives of the estimated partition function, e.g.,
$a_0 = \tilde E(0) = -\partial \ln Z/\partial \beta \big|_{\beta=0}$
and $a_1 =
\partial \tilde E/\partial \beta \big|_{\beta=0} = -\partial^2 \ln
Z/\partial \beta^2 \big|_{\beta=0}$... .

The simulation procedure is described as the follows. To be more
specific, we assume that a third order expansion of $\tilde
E(\beta)=a_0 + a_1\beta + a_2 \beta^2$ is used.  The method has two
components: a regular configurational sampling at a given
temperature $\beta$, and a random walk in the temperature space. For
generating configurational moves at a temperature $\beta$, the
Metropolis algorithm or a constant temperature molecular dynamics
method can be used.
After a configurational step is finished, we accumulate averages for
$\avl\beta\avr$, $\avl\beta^2\avr$, $\avl\beta^3\avr$,
$\avl\beta^4\avr$, $\avl E\avr$, $\avl \beta E\avr$ and $\avl
\beta^2 E\avr$, where the first four values correspond to $\avl
\phi_j(\lambda) \  \phi_i(\lambda) \wavr$ and the rest to $\Avl
\phi_j(\lambda) \  \avl P_\lambda(X)\lavr \  \wAvr$ in Eq.
(\ref{matrix}).
Note, the symbol $\avl\ldots \avr$ here is a shorthand notation for
a trajectory average; it corresponds to an ensemble average in the
generalized ensemble, where the averaging over both configuration
and  $\beta$ is implied.
For simplicity, we have also assumed $\tilde w(\beta)$ to be a
constant and thus dropped the $\tilde w$ subscript.

The temperature space random walk is realized by assuming $\beta$ as
a continuous variable within the temperature range of interest
$(\beta_0, \beta_1)$. Although one can divide the whole temperature
range into several sub-windows, we assume only one window in this
example for the sake of simplicity.  The current temperature $\beta$
is updated regularly, i.e., every a few configurational sampling
steps. Before a temperature update, the coefficients $a_0$, $a_1$
and $a_2$ are determined by solving Eq. (\ref{matrix}), or
explicitly
\[
\left( \begin{array}{ccc} %
1  & \avl \beta \avr  & \avl \beta^2 \avr \\
\avl \beta \avr  & \avl \beta^2 \avr  & \avl \beta^3 \avr \\
\avl \beta^2 \avr  & \avl \beta^3 \avr  & \avl \beta^4 \avr %
\end{array} \right) %
\left( \begin{array}{c} %
a_0    \\
a_1  \\
a_2  %
\end{array} \right) %
=
\left( \begin{array}{c} %
 \avl E \avr   \\
\avl \beta E\avr  \\
\avl \beta^2 E \avr  %
\end{array} \right). %
\]
Then the temperature $\beta$ is updated according to the Langevin
equation Eq. (\ref{lang}), or explicitly
\[
    d\beta/dt
    = \bigl(a_0+a_1\beta + a_2\beta^2)-E +\xi,
\]
where $E$ is the current energy; $\xi$ is a Gaussian white noise
that satisfies $\langle \xi(t) \xi(t') \rangle = 2 \delta(t-t')$,
which can be conveniently generated using a random number generator.
If the Langevin equation drives the current temperature out of the
entire temperature range, the update is rejected and the old
temperature is preserved.
At the end, the ratio of the estimated partition function can be
calculated as,
\begin{equation*}
\begin{split}
    & \ln \tilde Z(\beta_0) - \ln \tilde Z(\beta_1) \\
    & = a_0 (\beta_1-\beta_0) +
    a_1  (\beta_1^2-\beta_0^2)/2 +a_2  (\beta_1^3-\beta_0^3)/3.
\end{split}
\end{equation*}
As the simulation progresses, the coefficients $a_0$, $a_1$ and
$a_2$ gradually converge to fixed values, and the ratio of the
estimated partition function asymptotically approaches the ratio of
the correct one $ \tilde Z(\beta_0)/ \tilde Z(\beta_1) \rightarrow
Z(\beta_0) / Z(\beta_1)$.

Note, in this method, although $a_k$'s correspond to different order
derivatives of the partition function, the determination of these
parameters by Eq. (\ref{matrix}) does not involve averages of
high-order moments of $P_\lambda(X)$ such as $\Avl [P_\lambda(X)]^2
\wAvr$ or $ \Avl [P_\lambda(X)]^3 \wAvr$, or averages of high-order
derivatives of $\Hr(X)$ such as $\Avl \partial^2 \Hr(X)/\partial
\lambda^2 \wAvr = -\Avl \partial P_\lambda(X)/\partial \lambda
\wAvr$ and $ \Avl
\partial^3 \Hr(X)/\partial \lambda^3 \wAvr = - \Avl \partial^2
P_\lambda(X)/\partial \lambda^2 \wAvr$. This is a desirable feature
because these high-order quantities are usually difficult to compute
or may not be well-defined. We naturally avoid these quantities by
using moments of the variable $\lambda$ instead. In the above
example, $a_0$, $a_1$ and $a_2$ are determined by averages, such as
$\avl \beta^3 \avr$ and $\avl \beta^2 E \avr$, but not high-order
moments of $E$, such as $\avl \beta E^3 \avr$ and $\avl \beta E^2
\avr$. This feature makes the updating more robust and the method
more applicable for a general $\lambda$.

\section{Numerical Results}
\subsection{Two-dimensional Ising model}
\begin{figure}[h]
  \begin{minipage}{.9 \linewidth}
    \begin{center}
        \includegraphics[angle=-90,width=  \linewidth]{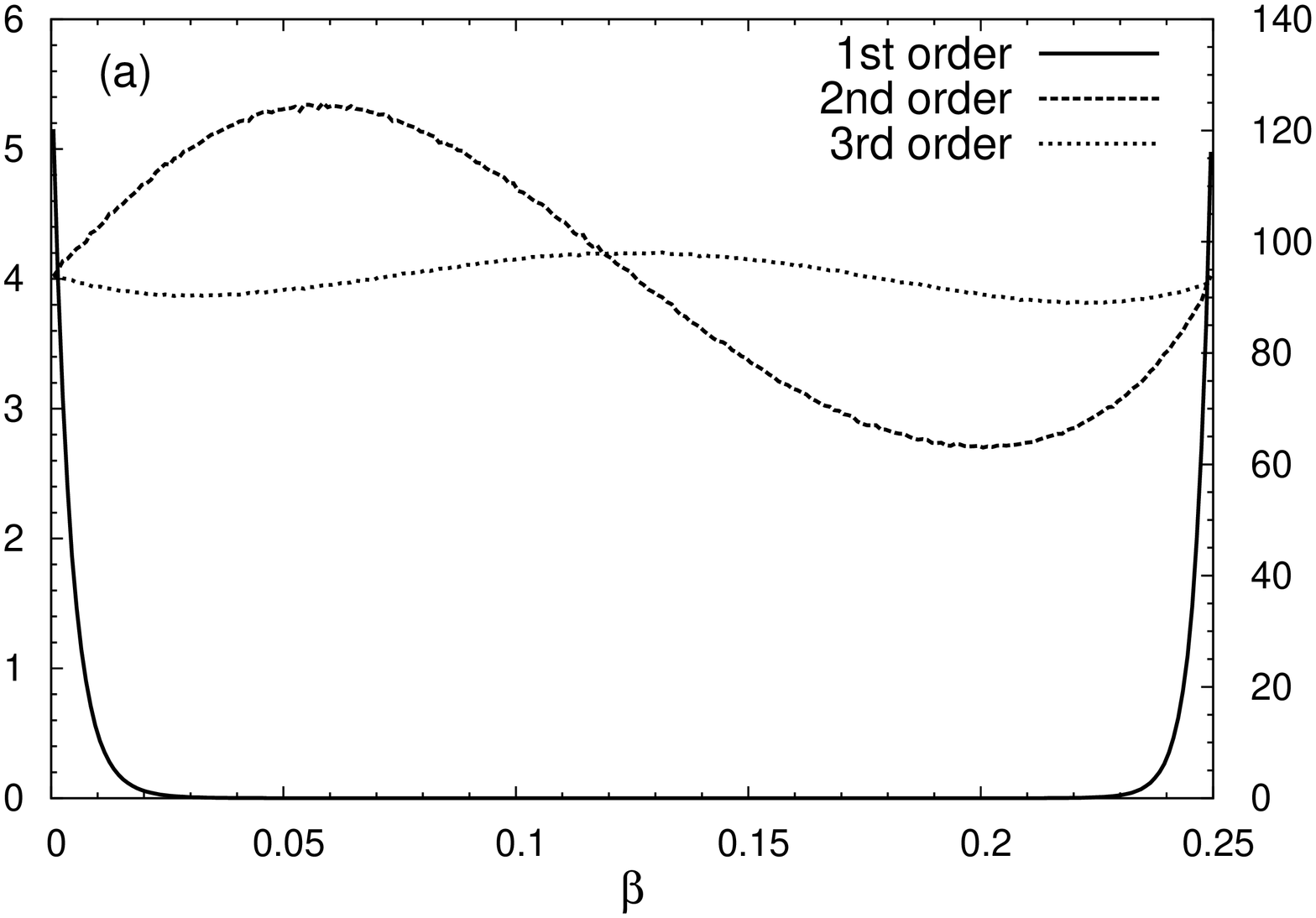}
    \end{center}
  \end{minipage} \\
  \begin{minipage}{.9 \linewidth}
    \begin{center}
        \includegraphics[angle=-90,width=  \linewidth]{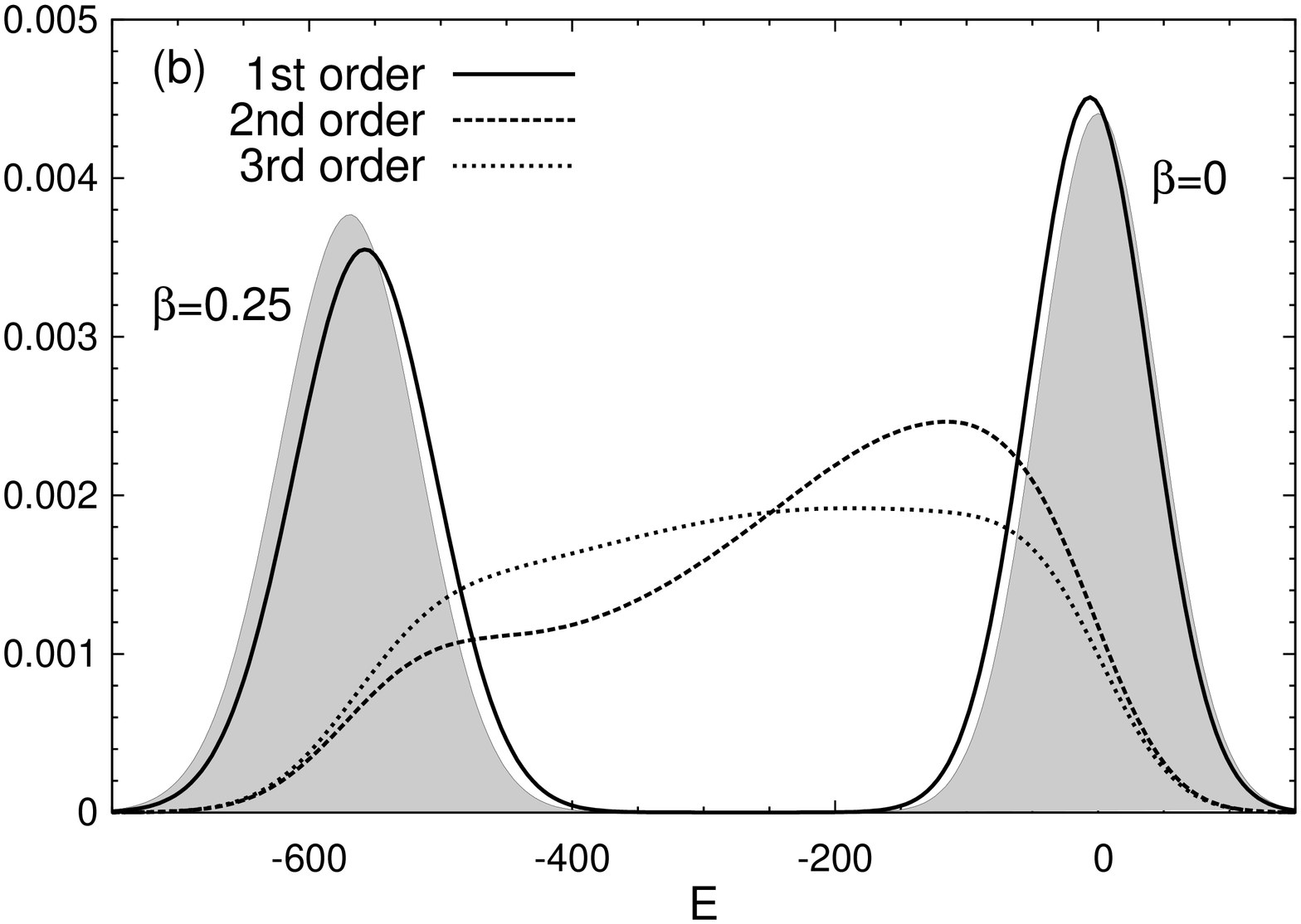}
    \end{center}
  \end{minipage}%
  \caption{\label{fig:hbf} %
  (a) The $\beta$-histogram using different orders of $\tilde E(\beta)$
  expansion.  The right axis is for the first order (the solid line).
  The left axis is for the rest two (the dashed and dotted line).
  (b) The corresponding energy histograms using different orders expansions.
  For comparison, the constant temperature energy histograms
   (using the Metropolis algorithm) for $\beta=0$
  and $\beta=0.25$ are shown in the shaded area.
  }
\end{figure}

We first perform a test on the $32\times 32$ Ising model using the
first, second, and third order series expansion of $\tilde
E(\beta)$.  The Metropolis algorithm is used to generate
configuration changes. The range of $\beta$ is $(0,0.25)$, the
corresponding $T$ range is $(4,+\infty)$.  The time step for
integrating the Langevin equation Eq. (\ref{lang}) is
$5\times10^{-5}$. The results are shown in Fig. \ref{fig:hbf}.
First let us examine the $\beta$-histogram, which corresponds to the
$\beta$-distribution $w(\beta)$ defined in Eq. (\ref{wZ}). According
to Eq. (\ref{convergence}), the values of the $\beta$-histogram at
the endpoints should be equal, i.e., $w(\beta_0) = w(\beta_1)$
[$\tilde w(\beta)$ is constant here and $Z(\beta_0)/\tilde
Z(\beta_0) = Z(\beta_1)/\tilde Z(\beta_1)$ according to Eq.
(\ref{convergence})]. Fig. \ref{fig:hbf}(a) agrees with the
expectation in all the three cases.
%
%
In addition, we expect the approximate partition function to be
sufficiently close to the exact one.  The difference of the two can
be examined from the difference between the actual weight $
w(\beta)$ and the desired weight $\tilde w(\beta)$. Since $\tilde
w(\beta)$ is a constant in this case, the $\beta$-histogram that
represents $w(\beta)$ should be sufficiently flat. It can be seen
from Fig. \ref{fig:hbf}(a), the first-order algorithm yields a
$\beta$-distribution peaked at both the boundaries $\beta=0$ and
$\beta=0.25$, while the third-order algorithm yields an almost flat
histogram.  For the energy histograms [Fig. \ref{fig:hbf}(b)], since
the corresponding energy distributions are very different at
$\beta=0$ and $\beta=0.25$, $\tilde E(\beta)$ cannot be represented
by a constant value, and thus the first order algorithm becomes
ineffective (no overlap between two energy distributions at the
given $\beta$-gap). On the other hand, using a higher order version,
$\tilde E(\beta)$ can more effectively approximate the average
energy as a function of $\beta$. As a result, we can achieve a
flatter $\beta$-histogram as well as a broader energy histogram. The
example shows that higher order algorithms can handle a much larger
temperature gap than the first order one.

We also compare the current method with the method from a previous
study \cite{Z} which uses the Wang-Landau (WL) updating scheme
\cite{wl} to converge the partition function.
In both cases, we update the temperature after a sweep of
configuration sampling. For the current method, the third order
expansion with a single temperature window is used.
For the previous method, twenty five sampling temperatures are
evenly distributed in the temperature range with $\Delta\beta=0.01$.
Such a fine temperature interval ensures that the previous method
targets the flat-$\beta$-histogram ensemble and it has a good
transition rate between neighboring temperatures. The parameters of
the WL updating scheme are the following. The initial value for the
modification factor $\ln f=1.0$ and it is shrunk by a factor of 2 at
the end of each stage. The criterion for terminating a stage depends
on the flatness of the temperature histogram. Three different
choices of the flatness thresholds 20\%, 50\% and 99\% are used (in
the last case, a stage is terminated when each sampling temperature
is visited at least once).
We also use a recipe \cite{1t} of improving the convergence of the
WL updating scheme in final stages, where the modification factor
$\ln f$ is specified as $1/t_n$ in final stages regardless of the
histogram flatness. Here $t_n$ is defined as the number of Monte
Carlo steps divided by the number of temperatures.

\begin{figure}[h]
  \begin{minipage}{ \linewidth}
    \begin{center}
        \includegraphics[angle=-90,width=  \linewidth]{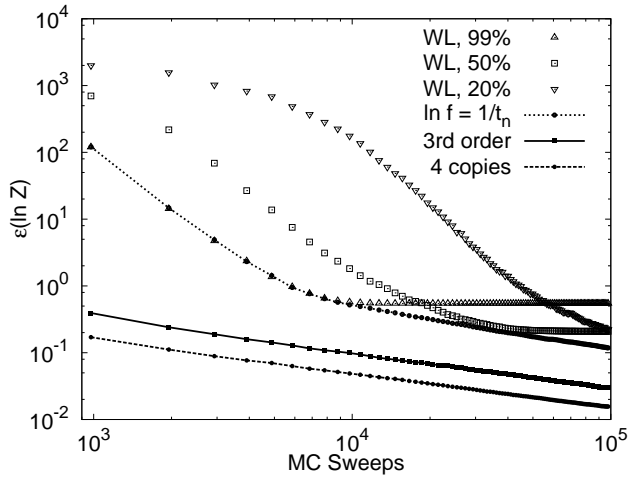}
    \end{center}
  \end{minipage}%
  \caption{\label{fig:perf} The logarithmic error of the ratio of the partition function
  $\epsilon \bigl( \ln [Z(\lambda_1)/\ln Z(\lambda_0)]\bigr)$ versus the simulation time $t$.
  For the WL updating
  scheme, three thresholds of the histogram flatness 20\%, 50\%, and 99\%
  are used.
  The result from the $\ln f = 1/t_n$ correction (where $t_n$ is the number of MC steps divided by
  the number of temperatures) is shown as the dotted line.
  For the current method (solid line), the error $\epsilon$ scales with the simulation time
  $t$ as $\epsilon \sim  t^{-0.52}$.  %
  By using the parallel version (four copies, dashed line), the error is further
  reduced.
  All results are averaged over 1000 independent runs.
  }
\end{figure}

The results of the comparison are shown in Fig. \ref{fig:perf}.
Since the exact partition function for the Ising model is available
\cite{isingexact}, the logarithmic error of the ratio of the
partition function, defined as $\epsilon = \bigl|\Delta \ln\tilde Z
- \Delta \ln Z\bigr|$, is used to measure the accuracy. For the
current method, the accuracy improves steadily as the simulation
progresses.  The error $\epsilon$ as a function simulation sweeps
(MC steps per site) $t$ can be fitted by regression as
$\epsilon=11.6 \ t^{-0.52}$.
It is clear that the original WL recursive scheme suffers from the
problem of saturation at a long simulation time. Although the
$1/t_n$ recipe eases the problem, it is still less efficient than
the method introduced here.
The error of the current method at the end of $10^5$ sweeps is
0.0297, while for the WL recursion with $1/t_n$ recipe, the error is
0.121.  The same accuracy $\epsilon = 0.121$ can be achieved by the
current method at the end of 6\ 000 sweeps.  This shows that the
current method is one to two orders of magnitude more efficient than
the WL recursion in terms of simulation time.

The efficiency of the method can be further improved when parallel
computers are available. We briefly describe a parallel extension
here.  In the parallel version, multiple copies of simulations run
simultaneously using a same set of $a_k$'s.  All copies contribute
to the trajectory averages, such as $\avl \beta^2\avr$ and $\avl
\beta E\avr$. The parameters $a_k$'s shared by all copies are
calculated from the averages from multiple trajectories and
therefore are more accurate. In Fig. \ref{fig:perf}, we also show
that the result from the parallel version (dashed line) using four
copies. The error at the end of $10^5$ sweeps is 0.0156, which is
about half of the single copy version.  According to the $t^{-1/2}$
scaling relation, the convergence rate is about four times as fast
as the single copy one, as we expected.  By contrast, the WL type
updating does not have a convenient parallel counterpart.

\subsection{Atomistic model protein}

The method is also applied to locate low energy states of a $AB$
model protein \cite{ab}, which has been extensively studied in
literature \cite{csa, st, Z}. It has two types of residues, $A$:
hydrophobic, $B$: hydrophilic. Particularly, we used the second set
of molecular force fields \cite{ab}, which produces more globular
low energy structures.

The major challenge of this system is that it contains many
different low-energy wells separated by high barriers. Due to its
rugged low-energy landscape, the $AB$ model serves as a stringent
testing case for the ability of configurational sampling of the
algorithm. Although in principle thermal properties are determined
by averages from all low-energy wells, only the one with the lowest
energy has a dominant contribution at a low temperature. For
example, if two energy wells have a $\Delta E=2.0$  difference in
their energy (which is common for a system with 55 or 89 residues),
the contribution from the higher energy well is only $\exp(-\Delta
E/T) \approx 2\times 10^{-9}$ times that of the lower energy well at
 temperature $T = 0.1$. Since lower energy states
are gradually discovered as the simulation proceeds, average
properties estimated in an earlier time must be promptly corrected
according to the newly found low energy states.

Due to this reason, we use a more aggressive averaging scheme that
favors recent statistics.
In this scheme, we introduce a memory factor $\gamma < 1$ to
gradually shrink the weight of previous
statistics.  %
For example, the average energy $\avl E\avr$ is computed as
$\mathcal{E/N}$, where the total energy $\mathcal E$  and the total
weight $\mathcal N$ (which are both accumulated from the beginning
of simulation) are updated as $\mathcal E \rightarrow \gamma
\mathcal E +E$ and $\mathcal N \rightarrow \gamma \mathcal N+1$,
where $\gamma\le1$.  If $\gamma=1$, the averaging scheme is reduced
to a regular average.  The factor $\gamma<1$ is particularly useful
in correcting low temperature statistics and in enhancing the
temperature-space random walk. However, $\gamma$ should still be
close to 1 to maintain a good sampling accuracy.  In this example,
$\gamma = 1.0 - 10^{-7}$ is used.

In implementation, Brownian dynamics is used for constant
temperature simulation.  The equation of motion is $d\vec x/d t =
\vec F +\vec \eta$ , where $\vec F$ is the force derived from a
molecular potential; $\vec \eta$ is a vector of Gaussian white noise
specified by $\langle \eta_i(t) \eta_j(t') \rangle = 2 T
\delta(t-t') \delta_{i,j}$; $T$ is the current temperature. The time
step is $3\times10^{-3}$.  For polymer of 34 and 55 residues, the
temperature range is $(0.1,0.7)$.  The method can easily locate the
known lowest energy configurations with $E = - 98.3571$ and
$-178.1339$ respectively \cite{Z}.  No other lower energy state is
discovered.

\begin{figure}[h]
  \begin{minipage}{ 0.5 \linewidth}
    \begin{center}
        \includegraphics[angle=-90, width=  \linewidth]{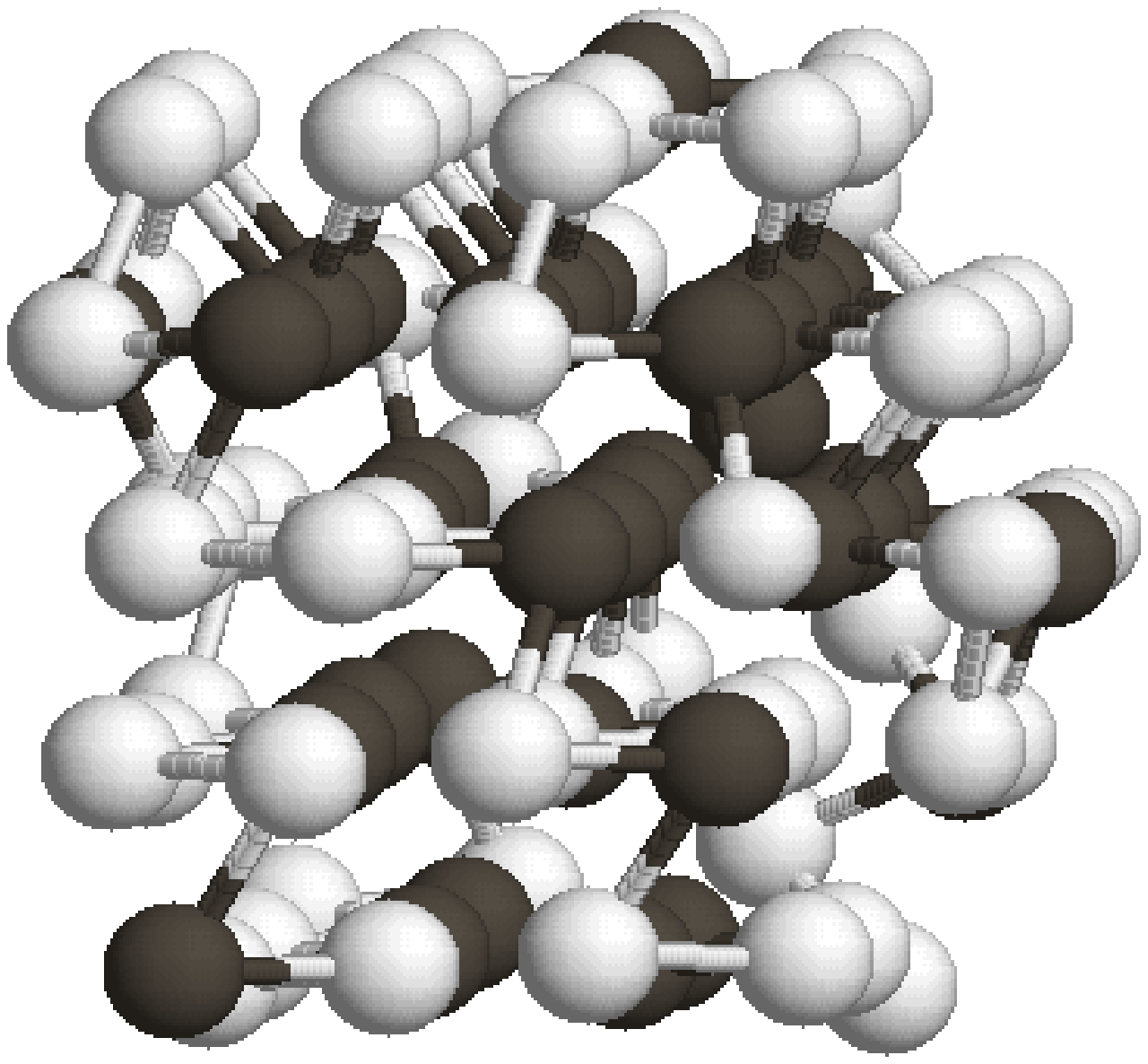}
    (a)
    \end{center}
  \end{minipage}%
  \begin{minipage}{ 0.5 \linewidth}
    \begin{center}
        \includegraphics[angle=-90, width=  \linewidth]{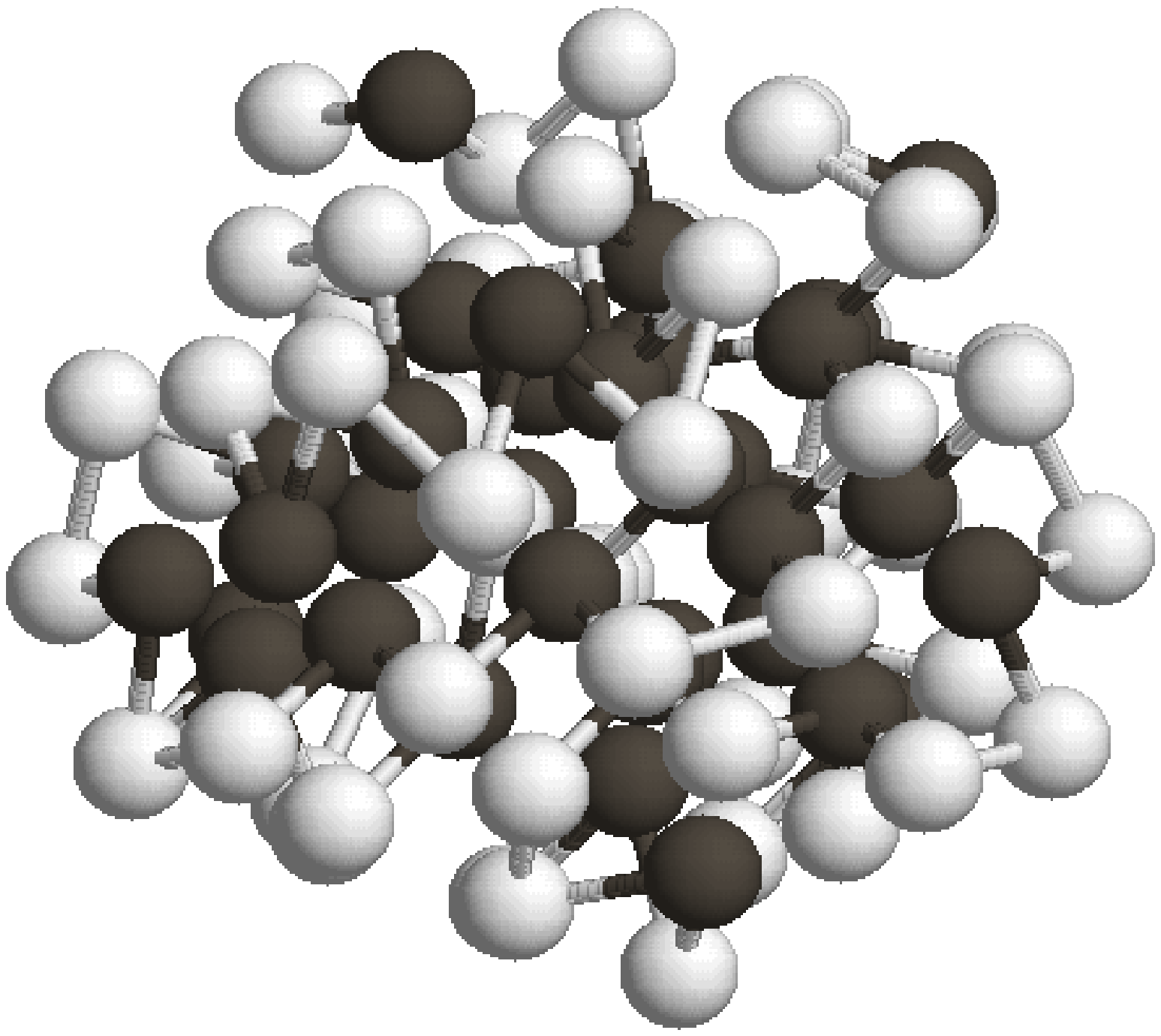}
    (b)
    \end{center}
  \end{minipage}%
  \caption{\label{fig:abcfg}
  Panel (a): the lowest-energy configuration of 89 residue model protein,
  $E=-311.6137$;
  Panel (b): a different configuration with a similar local minimal energy $E=-311.5391$
  (black, $A$; white, $B$).}
\end{figure}

The polymer of 89 residues is more challenging and its lowest energy
configuration has not been reported in literature to authors' best
knowledge. In this case, the temperature range is $T = (0.15, 0.4)$
with five windows, separated at $T=0.20$, $0.23$, $0.26$, and $0.3$.
Within each window, a second order expansion in term of $T$, $\tilde
E = a_0 + a_1 T$ is used. This expansion is more suitable than the
expansion in terms of $\beta$ at a low temperature because $a_0$ is
roughly the ground state energy while $a_1$ is the average heat
capacity. The weighting factor $\tilde w(T)\ dT \sim
1/\left\{1+[(T-T_c)/\Delta]^2\right\}\ dT$ concentrated on $T_c =
0.25$ with a width $\Delta=0.1$ is used to accelerate the
temperature-space random walk. In addition, since the goal is to
find the ground state instead of calculating the free energy, we
replaced the weighted averages by regular averages in Eq.
(\ref{matrix}) and remove the constraint Eq. (\ref{convergence}) to
make the averaging and updating process more stable. The lowest
energy state found in this study is $E = -311.6134$. The
corresponding configuration is shown in Fig. \ref{fig:abcfg}(a).
Here, another state with a very close energy $E = -311.5391$ but
with a very different configuration is also shown in Fig.
\ref{fig:abcfg}(b). The result indicates an extreme ruggedness of
the low energy landscape.

\section{Concluding Discussions}

In summary, we demonstrated an enhanced sampling method using a
generalized ensemble. The method computes the partition function by
minimizing the difference between the derivative of the estimated
partition function and that of the actual one. One advantage of the
method is that it allows a large gap of the macroscopic variable
$\lambda$ of the partition function.  For example, when
$\lambda=\beta$, it can afford a much larger temperature gap than
other tempering methods \cite{replica, simtemp, Z} that rely on
overlap between distributions.  This feature makes the method more
suitable for handling larger systems with much narrower
distributions of thermodynamic quantities.  The method also delivers
asymptotic convergence of the partition function, which makes it
superior to other methods based on the WL recursive scheme
\cite{wl}.  The efficiency of the method is demonstrated on the two
dimensional Ising model and the off-lattice protein models.

One of the most important features of our method is its scalability
to large systems.  The method performs a random walk in the
$\lambda$ space, e.g., the reciprocal temperature $\beta$ as in the
examples, and the ratio of partition functions between two
end-points of a $\lambda$ window is calculated by minimizing the
difference between the estimated and measured (from simulation
trajectory) values of the thermal conjugates of $\lambda$ (the
energy
in the case of $\lambda=\beta$). %
By including parameters that correspond to higher-order derivatives
of the partition function in the Langevin equation controlling the
$\lambda$-space random walk, the profile of the partition function
within the $\lambda$ window is more accurately approximated and the
Brownian motion in the $\lambda$ space is augmented.
Thus, as long as the thermal conjugates varies smoothly within a
$\lambda$ window, the method can bring an efficient sampling of the
entire $\lambda$ window. Such a feature makes the method much less
sensitive to the size of system, as well as the size of $\lambda$
window.

In this study, we demonstrated the efficiency of the method in the
reciprocal temperature space ($\lambda=\beta$).  However, $\lambda$
can be other variables.  In our previous study \cite{Z}, we used the
volume.  Besides, it can be other $\lambda$-parameter commonly used
in free energy simulation \cite{adaptive, ti, sscaling,
otherlambda}.

Strictly speaking, the current method does not satisfy detailed
balance due to the use of runtime averages, as in other algorithms
before convergence \cite{Z, wl, 1t, adaptive}. However, as
simulation progresses the correction to the existing averages
continuously decreases, and the deviation from detailed balance is
negligible in the asymptotic limit.  On the other hand, the runtime
averaging process is essential to continuously improve the estimate
of the partition function.


\section*{Acknowledgements}

The authors acknowledge support of grants from the National
Institutes of Health (R01-GM067801), the National Science Foundation
(MCB-0818353), and the Welch Foundation (Q-1512).


\begin{thebibliography}{widest-label}


\bibitem{Z}
    C. Zhang and J. Ma,
    Phys. Rev. E
    \textbf{76}, 036708 (2007);
    Phys. Rev. E
    \textbf{79}, 016703 (2009).
    J. Chem. Phys.
    \textbf{129}, 134112 (2008).


\bibitem{replica}
R. H. Swendsen and J. S. Wang,
    Phys. Rev. Lett. \textbf{57}, 2607 (1986);
C. J. Geyer, Proceedings of the 23rd symposium on the interface
    (American Statistical Association, New York, 1991); %
K. Hukushima and K. Nemoto, J. Phys. Soc. Jpn.
    \textbf{65}, 1604 (1996); %
U. H. E. Hansmann, Chem. Phys. Lett.
    \textbf{281}, 140 (1997).


\bibitem{simtemp}
A. P. Lyubartsev, A. A. Martsinovski, S. V. Shevkunov, and P. N.
Vorontsov-Velyaminov,
    J. Chem. Phys. \textbf{96}, 1776 (1991);
E. Marinari and G. Parisi,
    Europhys. Lett. \textbf{19}, 451 (1992).


\bibitem{sscaling}
H. Li, M. Fajer, and W. Yang,
    J. Chem. Phys. \textbf{126}, 024106 (2007);
    \textbf{129}, 034105 (2008).


\bibitem{adaptive}
E. Darve and A. Pohorille
    J. Chem. Phys. \textbf{115}, 9169 (2001);
M. Fasnacht, R.H. Swendsen, and J.M. Rosenberg,
    Phys. Rev. E. \textbf{69}, 056704 (2004).

\bibitem{wl}
F. Wang and D. P. Landau,
    Phys. Rev. Lett.
    \textbf{86}, 2050 (2001).
    Phys. Rev. E
    \textbf{64}, 056101 (2001).


\bibitem{1t}
R. E. Belardinelli and V. D. Pereyra,
    Phys. Rev. E \textbf{75}, 046701 (2007).

\bibitem{isingexact}
    A. E. Ferdinand and M. E. Fisher,
    Phys. Rev. \textbf{185}, 832 (1969).

\bibitem{ab}
    F. H. Stillinger, T. Head-Gordon, and C. L. Hirshfeld,
        Phys. Rev. E \textbf{48}, 1469 (1993);
    A. Irb\"{a}ck, C. Peterson, F. Potthast, and O. Sommelius,
        J. Chem. Phys. \textbf{107}, 273 (1997).

\bibitem{csa}
    S. Y. Kim, S. B. Lee, and J. Lee,
        Phys. Rev. E \textbf{72}, 011916 (2005);
    J. Lee, K. Joo, S. Y. Kim,  and J. Lee
        J. Comput. Chem. \textbf{29}, 2479 (2008).

\bibitem{st}
    J. G. Kim, J. E. Straub, and T. Keyes,
        Phys. Rev. Lett. \textbf{97}, 050601 (2006);
        Phys. Rev. E \textbf{76}, 011913 (2007).

\bibitem{ti}
    J. G. Kirkwood,
    J. Chem. Phys.
    \textbf{3}, 300 (1935).
    Carter EA, Ciccotti G, Hynes JT, Kapral R,
    Chem. Phys. Lett.
    \textbf{156}, 472 (1989).

\bibitem{otherlambda}
    G. M. Torrie and J. P. Valleau,
    J. Comput. Phys.
    \textbf{23}, 187 (1977);
    C. Bartels C, M. Karplus,
    J. Comput. Chem.
    \textbf{18}, 1450 (1997);
%
    P. A. Bash, U. C. Singh, R. Langridge, and P. A. Kollman
    Science
    \textbf{236}, 564 (1987);
%
    R. W. Zwanzig,
    J. Chem. Phys.
    \textbf{22}, 1420 (1954).
    D. J. Tobias and C. L. Brooks III,
    Chem. Phys. Lett.
    \textbf{142}, 472 (1987);
%
    X. J. Kong and C. L. Brooks,
    J. Chem. Phys.
    \textbf{105}, 2414 (1996);
%
    A. Laio, M. Parrinello,
    Proc. Natl. Acad. Sci. USA
    \textbf{99}, 12562 (2002);
%
    L. Zheng, M. Chen, and W. Yang,
    Proc. Natl. Acad. Sci. USA
    \textbf{105}, 20227 (2008);
%
    T. P. Straatsma, J. A. McCammon
    Ann. Rev. Phys. Chem.
    \textbf{43}, 407 (1992).

\end{thebibliography}
\end{document}